\documentclass[reprint,aps]{revtex4-1}
\usepackage[british]{babel}
\usepackage{amsmath,amssymb,graphicx,hyperref,xcolor,theorem}

\newcommand{\tmmathbf}[1]{\ensuremath{\boldsymbol{#1}}}
\newcommand{\tmop}[1]{\ensuremath{\operatorname{#1}}}

\newcommand{\tmtextit}[1]{\text{{\itshape{#1}}}}
\newcounter{nnacknowledgments}

{\theorembodyfont{\rmfamily}\newtheorem{acknowledgments*}[nnacknowledgments]{Acknowledgments}}

\begin{document}

\title{Travelling wave amplification in stationary gratings}

\author{ S. A. R. Horsley$^1$ and J. B. Pendry$^2$}
\affiliation{$^1$School of Physics and Astronomy, University of Exeter, Stocker Road, EX4 4QL}
\affiliation{$^2$The Blackett Laboratory, Department of Physics, Imperial College London, London, SW7 2AZ}

\

\begin{abstract}
We show that a grating amplitude stationary in space but oscillating in time can be accurately modelled as a set of independent gratings travelling in opposite directions, interacting almost exclusively with waves travelling in the same direction. This structure reproduces the key features of travelling gratings: amplification of a wave at points where the local wave speed equals the grating velocity. The same field compression and photon production is evident when even a single Fourier component of the grating profile has a velocity that matches the local wave speed. We speculate that these stationary but oscillating gratings may prove easier to realise experimentally than travelling gratings.
\end{abstract}

{\maketitle}

Impressive experimental progress has been made in modulating the permittivity of materials on time scales comparable to the period of radiation {\cite{broderick1997,galiffi2022,lustig2023,tirole2023,moussa2023}}. Even more challenging is modulation in space and time, creating structures, gratings for example, that appear to move with velocities comparable to the speed of light.  This is a more difficult experiment and the present paper proposes a compromise of a static grating but one whose amplitude is modulated in time.  We shall show that the new structure reproduces many of the key features of a moving grating.

Motivation for these experiments is the rich body of theoretical work on time dependent structures, which can be traced back to the early work of Morgenthaler {\cite{morgenthaler1958}}. Amongst the many effects to be seen are the breaking of time reversal symmetry {\cite{sounas2017}}, gain effects which transform uniform incident waves into a series of narrow intense pulses {\cite{pendry2021a,pendry2021b}}, generation of spontaneous radiation through novel quantum processes {\cite{horsley2023}}, and more (summarised in a recent review {\cite{galiffi2022}}).

Theoretical studies show that in moving gratings, interaction with radiation is particularly strong for waves in the same direction as the grating, particularly as the grating speed approaches that of the radiation. On the other hand radiation travelling in the opposite direction is hardly affected at all by the grating, unless the medium is very strongly modulated.  This gives rise to the following notion: why not take two gratings moving in
opposite directions? Each primarily addresses co--propagating radiation, largely ignoring opposed radiation, and we might expect the two gratings should in essence act independently. In principle two gratings with space-time reciprocal lattice vectors of $g = 2 \pi / L$ and $\Omega = 2 \pi / T$ should constitute a 2D grating. If our speculation is correct, the 2D nature factorises approximately into two, and we get two 1D experiments for the price of one! Furthermore if the two oppositely travelling gratings are of the same amplitude and differ only in the direction of travel, as illustrated by Eq.(\ref{eq:epsilon-example}),
\begin{equation}
  \begin{array}{cl}
    \varepsilon (x, t) & = 1 + \alpha [\cos (g x - \Omega t) + \cos (g x + \Omega t)]\\
    & \\
    & = 1 + 2 \alpha \cos (g x) \cos (\Omega t) 
  \end{array}\label{eq:epsilon-example}
\end{equation}
they interfere to form a spatial structure whose amplitude is modulated in time. Spatial modulation of radiation pumping the grating is relatively simple to achieve either through masking of the radiation or having the modulated material pre--structured. so that its non-linear response is periodic.

Another advantage of the configuration is that its response is symmetric in space so that suitably positioned mirrors at either end of the grating mimic an infinite grating, even if the finite section of grating is only one period in length.  In the following sections we solve an idealised model of a stationary grating.  Our calculations verify our speculations and validate the new system for possible experimental realisation.

We consider electromagnetic waves propagating along the $x$--axis, with a $y$ polarized electric field. In terms of the magnetic vector potential $\tmmathbf{A} = A \tmmathbf{e}_y$, Maxwell's equations for the electric $\tmmathbf{E} = - \partial_t A\tmmathbf{e}_y$ and magnetic $\tmmathbf{H} = (\mu_0 \mu)^{- 1} \partial_x A\tmmathbf{e}_z$ fields reduce to the following second order wave equation,
\begin{equation}
    \frac{\partial}{\partial x} \left( \frac{1}{\mu} \frac{\partial A}{\partial x} \right) - \frac{1}{c^2_0} \frac{\partial}{\partial t} \left( \varepsilon\frac{\partial A}{\partial t} \right) = 0. \label{eq:wave-equation}
\end{equation}
To simplify the problem we neglect absorption, dissipation, and reflections, assuming impedance matching where, $\varepsilon = \mu = n$, with $n (x, t)$ is the space--time dependent refractive index.  With these assumptions, Eq. (\ref{eq:wave-equation}) can be factorized into a pair of first order operators
\begin{equation}
    \left( \frac{\partial}{\partial x} \frac{1}{n} - \frac{1}{c_0}\frac{\partial}{\partial t} \right) \left( \frac{\partial}{\partial x} + \frac{n}{c_0} \frac{\partial}{\partial t} \right) A = 0
\end{equation}
indicating that left and right going wave motion is uncoupled, with waves moving according to the first order equations
\begin{equation}
  \frac{\partial A}{\partial x} \pm \frac{n}{c_0} \frac{\partial A}{\partial
  t} = 0,\label{eq:right-left-propagation}
\end{equation}
the plus sign holding for right--going waves, and the minus sign for left--going ones.  Solutions to Eq. (\ref{eq:right-left-propagation}) can be found using the method of characteristics {\cite{arfken2011}}.
%
%
\begin{figure}[h]
  \raisebox{0.0\height}{\includegraphics[width=8.01575823166732cm,height=4.25183654729109cm]{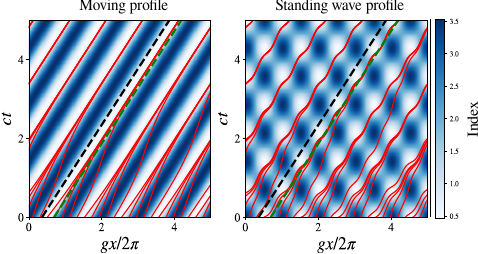}}
  \caption{\textbf{Concentration and rarefaction of waves in space--time gratings.} Rays (shown red) move with a local speed $c_0 / n$, becoming concentrated or rarefied along the green and black dashed diagonal lines respectively.  The above two panels show very similar effects occurring in (a) a travelling wave grating, $n (x, t) = 2.0 + 0.8 \cos (g (x - V t))$ and (b) a standing wave grating $n (x, t) = 2.0 + 0.3 \sin (g x) \cos (g V t) + 1.5 \cos (g x) \cos (g V t) .$\quad Here $V =
  0.7 c$.\label{fig:space-time-grating}}
\end{figure}

\begin{figure}[h]
  \raisebox{0.0\height}{\includegraphics[width=8.01575823166732cm,height=7.56268857405221cm]{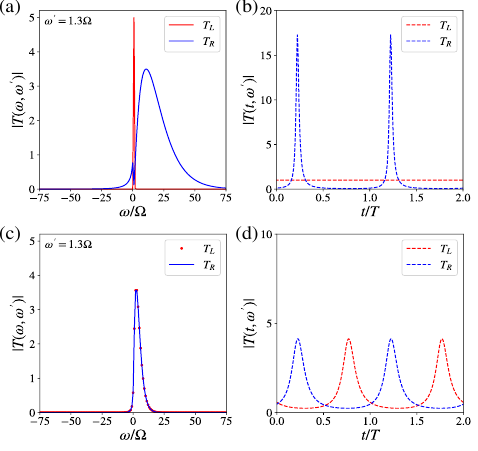}}
  \caption{\textbf{Transmission operator}.  Transmission of a monochromatic incident electric field of frequency $\omega' = 1.3 \Omega$, propagating either to the right (blue) and left (red). Panels (a--b) show the time and frequency dependence of the output for the travelling wave grating profile $n (x, t) = [1 + 0.1 \cos (g (x - Vt))]^{- 1}$, where $V =\Omega / g$.  Panels (c--d) show the same output for the standing--wave profile $n (x, t) = [1 + 0.1 \cos (g x) \cos (\Omega t))]^{- 1}$.  Simulation parameters are $c_0 = 1$, $T = 1$, $L = 1$, grating length is $4.5 L$, and simulation time is $5 T.$ \label{fig:transmission-function}}
\end{figure}

As shown in {\cite{pendry2021b,pendry2021a}}, a moving space--time grating can amplify electromagnetic waves, either concentrating or rarefying the energy at points where the local wave velocity equals the grating velocity.  An accumulation of energy occurs at one of these points, provided the wave moves more slowly immediately in front of this point, and faster immediately behind it.  Nearby wave energy is then forever funneled into this point, leading to a continual increase in the energy density during propagation.  This effect is demonstrated in the ray tracing calculation shown in Fig. \ref{fig:space-time-grating}a.  Fig. \ref{fig:space-time-grating}b, shows a similar effect also occurs in time varying index profiles where there is no identifiable motion at all.  In contrast to a moving grating, the profile in Fig. \ref{fig:space-time-grating}b is invariant under time reversal $t \rightarrow - t$.  Yet there remain the same lines in space time where wave energy is either concentrated or extinguished.  Due to the time reversal symmetry of the grating profile, these lines of energy accumulation and rarefaction interchange roles when the direction of wave propagation is reversed (see Fig. \ref{fig:transmission-function}d).

We can predict the accumulation of wave energy in these index profiles through finding the motion of an ultra--short pulse along a trajectory $x(t)$.  The local velocity of such a right--moving pulse is $+ c_0/ n$, i.e.
\begin{equation}
  \frac{\text{d} x (t)}{\text{d} t} = \frac{c_0}{n (x (t), t)} = \langle c\rangle + \eta (x (t), t) \label{eq:dxdt}
\end{equation}
where we have split the local velocity into the average speed along the trajectory $\langle c \rangle$, plus the deviation $\eta$.  Similarly writing the pulse motion $x (t)$ in the form of the average trajectory plus a fluctuation $\xi (t)$ we have
\begin{equation}
  x (t) = x_0 + \langle c \rangle t + \xi (t) .\label{eq:trajectory}.
\end{equation}
In terms of these variables, to leading order the equation of motion (\ref{eq:dxdt}) becomes
\begin{equation}
  \frac{\text{d} \xi}{\text{d} t} = \eta + \xi  \frac{\partial
  \eta}{\partial x_0} \label{eq:leading-order}
\end{equation}
where $\eta$ is evaluated along the trajectory $x_0 + \langle c\rangle t$.  It is now simple to integrate Eq. (\ref{eq:leading-order}), giving the solution
\begin{equation}
  \xi (t) = \xi (0) \exp \left( \int_0^t \frac{\partial \eta (t')}{\partial
  x_0} \text{d} t' \right) + \xi_1 (t) \label{eq:fluctuation}
\end{equation}
where we have defined
\begin{equation}
  \xi_1 (t) = \int_0^t \eta (t') \exp \left( \int_{t'}^t \frac{\partial \eta
  (t'')}{\partial x_0} \text{d} t'' \right)  \text{d} t' .
\end{equation}
The initial displacement $\xi (0)$ away from the line $x_0 + \langle c \rangle t$ thus exponentially grows or decays depending on the sign of the average slope of the wave speed $\partial \eta / \partial x_0$, defined in (\ref{eq:dxdt}).  Thus, just as for the truly moving profiles discussed in {\cite{pendry2021b,pendry2021a}}, all initial positions of the wave exponentially approach the same trajectory $x_0 + \langle c \rangle t + \xi_1 (t)$, if, in the vicinity of this path the \textit{average} wave speed \textit{decreases} with increasing $x$.  This explains the concentration of rays along the oscillating paths shown in Fig. \ref{fig:space-time-grating}b, showing that the amplification effect identified in {\cite{pendry2021b,pendry2021a}} is not unique to travelling wave gratings.  The wave is concentrated by the grating components that are co--propagating with the wave, the remaining grating components causing the oscillation of the wave as it is focused.

\begin{figure}[h]
  \raisebox{0.0\height}{\includegraphics[width=8.01575823166732cm,height=3.45026564344746cm]{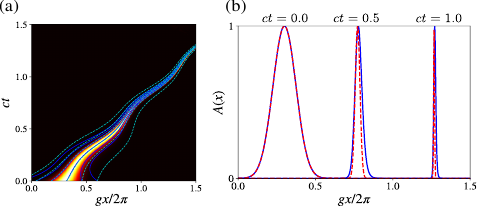}}
  \caption{\textbf{Wave compression in a standing wave grating.} Taking the index profile $n (x, t) = [1 + 0.8 \cos (g x) \cos (\Omega t)]^{- 1}$, panel (a) shows a color map of the approximate wave packet solution (\ref{eq:gaussian-approx}), overlaid with the approximation (\ref{eq:trajectory}) (solid) and exact integration (dashed) of the ray equation (\ref{eq:dxdt}).  Panel (b) shows a comparison between the wave packet solution to the wave equation
  (\ref{eq:gaussian-approx}) (red dashed) and a numerical integration of the equation (blue solid) for three sampled times during the propagation.  Parameters are $T = 1$, $L = 1$, $\sigma (0) = 0.074$, $\xi (0) = 0.3$, and $c_0 = 1$.\label{fig:approx}}
\end{figure}

To verify these approximate results, Fig. \ref{fig:approx}a shows that the trajectory (\ref{eq:trajectory}) with the approximate evolution (\ref{eq:fluctuation}) closely reproduces the results from a direct numerical integration of Eq. (\ref{eq:dxdt}).
%
%

To determine these lines along which wave energy becomes concentrated, we still must find the average wave velocity $\langle c \rangle$ introduced in Eq. (\ref{eq:dxdt}) which so far has remained undetermined.  To do this we expand the space--time varying wave velocity as a Fourier series with Fourier coefficients $\alpha_{p,q}$,
\begin{equation}
  c (x, t) = \sum_{p, q} \alpha_{p, q} \text{e}^{\text{i} (p g x - q \Omega t)} \label{eq:speed-expansion}
\end{equation}
Time averaging both sides of Eq. (\ref{eq:speed-expansion}) along the trajectory $x_0 + \langle c \rangle t$, only the part of the Fourier sum in (\ref{eq:speed-expansion}) which moves at velocity $\langle c \rangle$ survives, i.e.
\begin{equation}
  \langle c \rangle - \sum_p \alpha_{p, q_p} \text{e}^{\text{i} p g x_0} = 0 \label{eq:average-c}
\end{equation}
where we require $q_p \Omega / p g = \langle c \rangle$.  For the wave compression/rarefaction effect to occur, we must thus find a velocity $\langle c \rangle$, such that the part of the grating moving at the same velocity also contains points $x_0$ where the wave speed equals $\langle c\rangle$.  This calculation gives the slope $\langle c \rangle$ and offset $x_0$ of the green and black dashed lines shown in Figs. \ref{fig:space-time-grating} and \ref{fig:approx}.

Exactly the same predictions arise from solutions to the full wave equation (\ref{eq:right-left-propagation}).  Transforming Eq. (\ref{eq:right-left-propagation}) for a right--going wave into a new set of coordinates $x = x' + x_0 + \langle c \rangle t$ we have, to leading order in the displacement $x'$,
\begin{equation}
  \left( \eta + x' \frac{\partial \eta}{\partial x_0} \right) \frac{\partial
  A}{\partial x'} + \frac{\partial A}{\partial t} = 0 \label{eq:expanded}
\end{equation}
where, as in Eq. (\ref{eq:leading-order}), $\eta$ is evaluated along the line $x = x_0 + \langle c \rangle t$.  The general solution to Eq. (\ref{eq:expanded}) can be written as a arbitrary function $f$ of a polynomial,
\begin{equation}
  A (x', t) = f \left(  \sum_n a_n (t) {x'}^n \right) \label{eq:ansatz}
\end{equation}
where the time dependent functions $a_n$ in the series obey the coupled first order differential equations $\text{d} a_n / \text{d} t + n \partial_{x_0}\eta a_n = - (n + 1) \eta a_{n + 1}$.  Taking the special case of an exponential wave packet $f$, the series in Eq. (\ref{eq:ansatz}) can be terminated after three terms, $a_2$, $a_1$, and $a_0$  The wave then takes the form of a Gaussian
\begin{equation}
  A (x', t) = \exp \left( - \frac{1}{2 \sigma^2} (x' - \xi)^2 \right)
  \label{eq:gaussian-approx}
\end{equation}
where the time dependent width $\sigma$ and central position $\xi$ are given by
\begin{equation}
  \begin{array}{cl}
    \sigma & = \sigma (0) \text{e}^{\int_0^t \text{d} t' \frac{\partial
    \eta}{\partial x_0}}\\
    & \\
    \xi & = \xi (0) \text{e}^{\int_0^t \text{d} t' \frac{\partial
    \eta}{\partial x_0}} + \int_0^t \text{d} t' \eta (t')
    \text{e}^{\int_{t'}^t \text{d} t'' \frac{\partial \eta}{\partial x_0}} .
  \end{array} \label{eq:gaussian-parameters}
\end{equation}
As anticipated, the centre of the right--moving Gaussian $x' = \xi (t)$ in Eq. (\ref{eq:gaussian-parameters}) moves along the same trajectory as found using the ray approximation (\ref{eq:fluctuation}).  The
time dependent width $\sigma$ shows the expected energy compression/dilution in the region around $x' = 0$, again depending on the sign of the average value of $\partial \eta / \partial x_0$.  Fig. \ref{fig:approx}b shows that a direct numerical integration of Eq. (\ref{eq:expanded}) reproduces the same wave packet dynamics.

As an independent verification we now apply the
semi--analytical method described in Ref. {\cite{horsley2023b}}, to construct the transmission operator $\mathcal{T} (t, t')$ connecting the electric field on the two sides of a finite (length $L_g$) time--varying grating, $E_R (t) = \int_{-\infty}^{\infty} \mathcal{T} (t, t') E_L (t') \text{d} t'$.  For an impedance matched medium, the transmission operator can be approximated as a product of operators $\exp( \text{i} \hat{K}_n \Delta x)$ for homogeneous slabs of width $\Delta x$, where $\hat{K}_n = ( \frac{\omega}{c} \hat{\varepsilon} \frac{\omega}{c} \hat{\mu} )^{1 / 2}$ is the wave--vector operator for the $n^{\tmop{th}}$ spatial layer of the medium, written in terms of the time--varying permittivity and permeability as $\hat{\varepsilon} = \hat{\mu} = \widehat{\mathcal{F}} n (x_n, t) \widehat{\mathcal{F }}^{- 1}$ where $\widehat{\mathcal{F}}$ is a Fourier transform operator.

Fig. \ref{fig:transmission-function} shows the results from applying this transmission operator to an incident plane wave, comparing the time and frequency spectra of the pulses outgoing from the grating. As anticipated from the ray trajectories shown in Fig. \ref{fig:space-time-grating}, both moving and standing wave gratings lead to compressed output pulses and spectral broadening.  An important difference is that a truly moving grating has a highly asymmetric transmission function: a wave moving co--linear with the grating as in Fig. \ref{fig:transmission-function}a--b (blue lines)
is compressed in time and significantly broadened in frequency, whereas a counter moving wave is almost unaffected (red lines).  Meanwhile, transmission through the standing wave grating shown in Fig. \ref{fig:transmission-function}c--d is almost identical for the two directions of incidence, arising from the balance of both left and right moving Fourier components in the grating profile.
%
%

In practice, the mirror-symmetrical nature of the grating lends itself to shortening to one period and applying mirrors to the ends.  Light then makes repeated passes through the grating section. The mirrors would need to be positioned correctly so that reflected pulses hit the grating at the right moment for amplification. This would produce a unique amplifier: it would take a plane wave input, and output a frequency comb in the form of a train of sharp intense pulses. Alternatively the spontaneous emission
discussed in the next section could be exploited to form a ``qubit'' laser: photons are generated in pairs of opposite spin and concentrated into the pulses shown in Fig. \ref{fig:transmission-function}.

Ref. {\cite{horsley2023}}, demonstrated that the intense pulse compression in the moving grating shown in Fig. \ref{fig:transmission-function}b also leads to the emission of photons from the vacuum state.  This process happens whenever the grating is able to convert a positive frequency incoming wave into a negative frequency outgoing one, or vice versa.  We now show that the same process occurs also for the standing wave gratings considered here.

As established in the above discussion, an ultra short pulse of the vector potential $A$ moves according to the equation of motion (\ref{eq:dxdt}).  The transmission function $\mathcal{T}_A$ for the vector potential across the entire grating from $x = 0$ to $x = L_g$ can therefore be written as,
\begin{equation}
  \mathcal{T}_A (t, t') = \left| \frac{\partial x_c}{\partial t'} \right|
  \delta (L - x_c (t, t'))
\end{equation}
where $x_c (t, t')$ is the solution to the equation of motion (\ref{eq:dxdt}) for an initial position $x = 0$ at $t'$, and the prefactor $| \partial x_c /
\partial t' |$ ensures $\mathcal{T}_A (t, t') = \delta (t - t')$ in the limiting case where $L_g = 0$.  A single Fourier component of this transmission operator is thus,
\begin{equation}
  \begin{array}{cl}
    \mathcal{T}_A (t, \omega') & = \int_{- \infty}^{\infty} \mathcal{T}_A (t,
    t') \text{e}^{- \text{i} \omega' t'} \text{d} t'\\
    & \\
    & = \text{e}^{- \text{i} \omega' t' (t)} \label{eq:TAtw}
  \end{array}
\end{equation}
where $t' (t)$ is the entrance time such that the pulse is at $x = L_g$ at the exit time $t$, i.e. it is defined by
\begin{equation}
  x_c (t, t' (t)) = L_g . \label{eq:tpdef}
\end{equation}
Eq. (\ref{eq:TAtw}) shows the grating effectively performs a transformation of the time coordinate between the input and output facets of the medium, $\exp\left( - \text{i} \omega' t \right) \rightarrow \exp \left( - \text{i} \omega'
t' (t) \right)$.  Blue shifted frequencies arise from a compression of the initial time coordinate, whereas red shifts arise from an expansion of this coordinate.  Interestingly, negative output frequencies are generated from a positive frequency input when the output time coordinate is inverted relative to the input one, reminiscent of the coordinate transformation underlying negative refraction {\cite{pendry2000,leonhardt2012}}.

We can make this intuitive relationship precise through performing a second Fourier transform,
\begin{equation}
  \begin{array}{cl}
    \mathcal{T}_A (\omega, \omega') & = \int_{- \infty}^{\infty} \mathcal{T}_A
    (t, \omega') \text{e}^{\text{i} \omega t} \text{d} t\\
    & \\
    & = \int_{- \infty}^{\infty} \text{e}^{\text{i} (\omega t - \omega' t'
    (t))} \text{d} t.
  \end{array} \label{eq:T-fourier}
\end{equation}
Approximating the integral using the method of steepest descent
{\cite{bender1999}}, its value is dominated by all points in time, $t_s$ where $\frac{\text{d} t' (t_s)}{\text{d} t_s} = \frac{\omega}{\omega'}$.  This validates the intuition that the frequency rescaling $\omega / \omega'$ is due an effective compression or rarefaction of the output time axis $t' (t)$, provided by the material's time variation.

\begin{figure}[h]
  \raisebox{0.0\height}{\includegraphics[width=8.01575823166732cm,height=7.84149940968123cm]{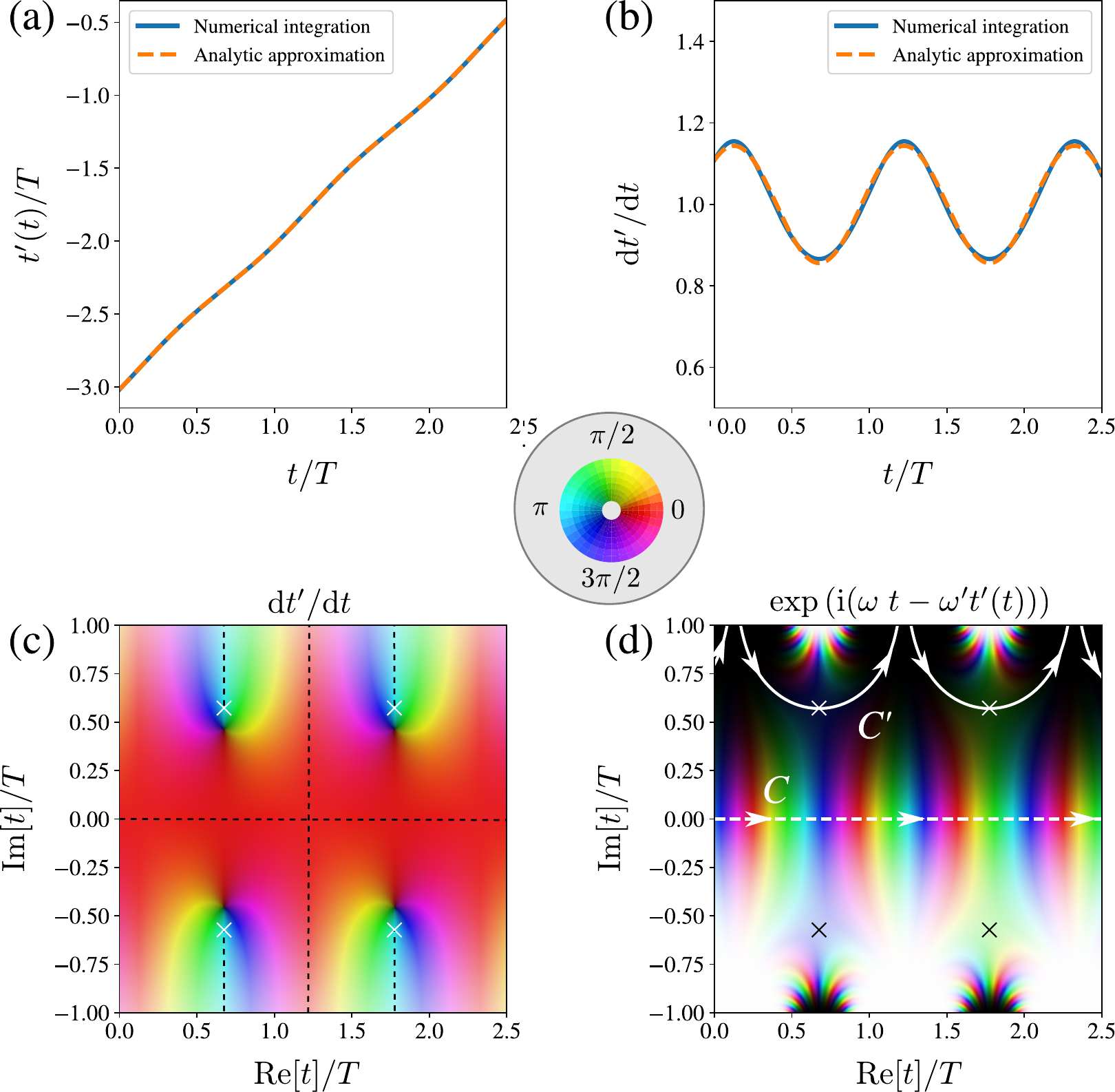}}
  \caption{\textbf{Negative frequencies and complex times:}  Taking a weakly modulated standing wave grating of the form
  (\ref{eq:epsilon-example}) with $\alpha = 0.01$, panels (a) and (b) show the entry time $t'$ and its derivative defined in (\ref{eq:tpdef}), comparing a direct integration of Eq. (\ref{eq:dxdt}) to the analytic result (\ref{eq:tp-approx}).\quad Panel (c) shows a phase plot of the derivative $\text{d} t' / \text{d} t$ (from Eq. (\ref{eq:tp-approx})) in the complex time plane, with dashed lines indicating negative real values.  Panel
  (d) shows a phase plot of the integrand (\ref{eq:T-fourier}), with the initial contour $C$ deformed into the series of loops $C'$ passing through saddle points (crosses).  Parameters are $\omega' = - 5$, $\omega = 4.5$, $T = 1.1$, $L = 1$, $\alpha = 0.01$, and $L_g = 3 L$.
  \label{fig:rescaling}}
\end{figure}

As the emission of photon pairs is governed by the part of the transmission function coupling positive and negative frequencies, it requires moments when the time axis is \tmtextit{reversed}, $\text{d} t' (t_s) / \text{d} t_s < 0$.  Meanwhile, as is evident in e.g. Fig. \ref{fig:rescaling}a, the arrival time $t' (t)$ is here a monotonically increasing function of the entry time, meaning that---as shown in Fig. \ref{fig:rescaling}b---$\text{d} t' (t) / \text{d} t$ is \tmtextit{positive} for all real $t$!  Therefore---assuming a suitable analytic continuation can be found---the coupling between positive and negative input/output frequencies is here governed by behaviour of the integrand of (\ref{eq:T-fourier}) at \tmtextit{complex} times.
%
%

Take for instance the space--time dependent permittivity and permeability given in Eq. (\ref{eq:epsilon-example}).\quad An approximation to the function $t' (t)$ can be found through directly integrating Eq. (\ref{eq:dxdt}),
\begin{equation}
  \begin{array}{ccl}
    \int_t^{t'} \text{d} t & = t' (t) - t & = \frac{1}{c_0} \int_L^0 n (x, t
    (x)) \text{d} x\\
    &  & \\
    &  & \sim \frac{1}{c_0} \int_L^0 n \left( x, t - \frac{L - x}{\langle c
    \rangle} \right) \text{d} x \label{eq:dtintegral}
  \end{array}
\end{equation}
where the final line neglects the fluctuation term in the trajectory (\ref{eq:trajectory}), valid for a weak modulation. For the particular profile (\ref{eq:epsilon-example}) this integral is,
\begin{equation}
  \begin{array}{c}
    t' (t) = t - \frac{L}{c_0} - \frac{\alpha a}{c_0} \cos (\Omega t - \psi) 
    \text{} \label{eq:tp-approx}
  \end{array}
\end{equation}
where $a e^{i \psi} = i \sum_{\pm} (e^{\mp i g L} - e^{i \Omega L / c_0})(\Omega / c_0 \pm g)^{- 1}$. Fig. \ref{fig:rescaling}a--c shows that the approximate expression (\ref{eq:tp-approx}) closely matches a full numerical integration of Eq. (\ref{eq:dtintegral}), and that $\text{d} t' / \text{d} t$ takes negative real values along lines in the complex time plane, parallel to the imaginary axis.

Considering only frequencies that change sign after propagation and deforming the integration contour $C$ in Eq. (\ref{eq:T-fourier}) from the real line to the series of loops $C'$ shown in Fig. \ref{fig:rescaling}d, the integrand of the Fourier transform is now exponentially small except close to the saddle points (identified as white crosses in Fig. \ref{fig:rescaling}c,d).  We thus
approximate the contour integration through each saddle point as a Gaussian integral.  The two frequency transmission operator (\ref{eq:T-fourier}) can then be approximated as
\begin{equation}
  \begin{array}{cl}
    \mathcal{T}_A (\omega, \omega') & = \tau_A (\omega, \omega') \sum_{n = -
    \infty}^{\infty} e^{i (\omega - \omega') n T}
    \label{eq:stationary-phase-T}\\
    & \\
    & = \tau_A (\omega, \omega') \frac{2 \pi}{T} \sum_{n = - \infty}^{\infty}
    \delta (\omega - \omega_n')
  \end{array}
\end{equation}
where $\omega_n' = \omega' + n \Omega$, and the single period transmission operator $\tau_A$ is defined as,
\begin{equation}
  \tau_A (\omega, \omega') = \sqrt{\frac{2 \pi}{\text{i} \omega'}
  \frac{1}{\frac{\text{d}^2 t' (t_s)}{\text{d} t_0^2}}} \text{e}^{\text{i}
  (\omega t_s - \omega' t' (t_s))} \label{eq:saddle-point-transmission}
\end{equation}
where the final sum over delta functions in Eq. (\ref{eq:stationary-phase-T}) arises from the Poisson summation formula {\cite{NIST:DLMF}}, enforcing the condition that the time periodic grating can only ever change the incident frequency in multiples of $\Omega$.  Note that in this case, we can also exactly evaluate the Fourier transform (\ref{eq:T-fourier}) in terms of Bessel functions (see Appendix).
%
%
\begin{figure}[h]
  \raisebox{0.0\height}{\includegraphics[width=8.01575823166732cm,height=2.7880919585465cm]{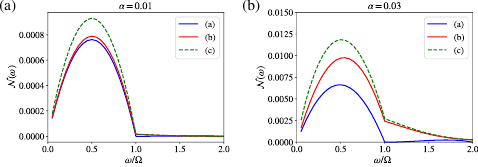}}
  \caption{\textbf{Photon emission from a standing wave grating}.  Number of photons per unit time, per unit frequency computed from Eq. (\ref{eq:photon-spectrum}), for two different amplitudes, $\alpha$ of the space--time varying grating (\ref{eq:epsilon-example}).  (a) Numerical calculation of the transmission operator (as in Fig. \ref{fig:transmission-function}); (b) the analytic calculation of the transmission function based on Bessel functions (see appendix); and (c) the saddle point approximation in the second line of (\ref{eq:photon-spectrum}).  As discussed in Ref. {\cite{horsley2023}}, numerical results show vanishing emission at frequencies that are multiples of the grating frequency $\Omega$.  This vanishing emission is not completely captured by the approximations (b) and (c).  Parameters are $T = 1.01$, $L = 1$, and $L_g = 4.5L$.\label{fig:photon-emission}}
\end{figure}

From this expression for the transmission function, the time averaged photon flux from the vacuum state can be found using the method given in {\cite{horsley2023}}.  We find---from Eq. (\ref{eq:saddle-point-transmission})---the average number of photons $\mathcal{N} (\omega)$ per unit frequency, per unit time, emerging from the grating is given by
\begin{equation}
  \begin{array}{cl}
    \mathcal{N} (\omega) & = \frac{1}{2 \pi T^2} \sum_{\omega_n < 0} \left(
    \frac{\omega}{- \omega_n} \right) | \tau_A (\omega, \omega_n) |^2\\
    & \\
    & = \frac{1}{T^2} \sum_{\omega_n < 0} \left( \frac{\omega}{\omega_n^2
    \left| \frac{\text{d}^2 t' (t_n)}{\text{d} t_0^2} \right|} \right)
    \text{e}^{- 2 \tmop{Im} [\omega t_n - \omega_n t' (t_n)]}
    \label{eq:photon-spectrum}
  \end{array}
\end{equation}
where $t_n$ is the saddle point defined by $\frac{\text{d} t' (t_n)}{\text{d}t_n} = \frac{\omega}{\omega_n}$.  Neglecting contributions from all saddle points but that nearest the real time axis, Eq. (\ref{eq:photon-spectrum}) shows that, as observed moving grating in {\cite{horsley2023}}, the number of emitted photons in the standing wave grating (\ref{eq:epsilon-example}) decays exponentially with $\omega$. Here we see that this is a rather generic behaviour where rate of decay is---to leading order---determined by the distance of the complex time where the grating time reverses the input wave, from the real time axis.

To conclude, we have shown that wave effects identified in moving periodic index profiles are inherited by non--moving standing wave gratings.  A standing wave grating can be decomposed into a set of counter--propagating travelling wave profiles, and those components where the local wave speed matches the grating speed lead to the same compression of the field predicted in a purely travelling grating.  The effect of the other grating components is to cause the wave to oscillate as it is compressed (see Figs. \ref{fig:space-time-grating} and \ref{fig:approx}), which also leads to an overall slower rate of compression (see Fig. \ref{fig:transmission-function}).

Another effect inherited from travelling wave index profiles is the emission of photons from the vacuum state.  Although the level of emission is---as expected---reduced compared to a purely travelling wave profile (compare Fig. \ref{fig:photon-emission} to Ref. {\cite{horsley2023}}), the form of the emission spectrum is largely the same.  We have also provided an approximate general formula (\ref{eq:photon-spectrum}) for the emission spectrum from a generic impedance matched space--time grating.  Positive to negative frequency conversion provided by the grating is due to time intervals where the medium time reverses the incident wave.  In most cases this occurs for complex values of time, leading to an overall exponentially decaying spectrum with frequency.  Yet, it would be interesting to explore space--time varying material profiles that maximize these intervals of time reversed transmission.

\begin{acknowledgments*}
SARH thanks Ian Hooper, James Capers, and Calvin Hooper for useful discussions and the Royal Society and TATA for financial support (URF\textbackslash R\textbackslash 211033).\quad JBP acknowledges funding from the Gordon and Betty Moore Foundation.
\end{acknowledgments*}

\section*{Appendix}

Here we directly evaluate the Fourier representation of the transmission function given in Eq. (\ref{eq:T-fourier}) using the approximate expression for the grating entry time (\ref{eq:tp-approx}),
\begin{equation}
  \begin{array}{cl}
    \mathcal{T}_A (\omega, \omega') & = \int_{- \infty}^{\infty}
    \text{e}^{\text{i} (\omega t - \omega' t' (t))} \text{d} t\\
    & \\
    & = \int_{- \infty}^{\infty} \text{e}^{\text{i} \left( (\omega - \omega')
    t + \frac{\omega' L}{c_0} + \frac{\alpha a \omega'}{c_0} \cos (\Omega t -
    \psi) \right)} \text{d} t\\
    & \\
    & = 2 \pi \sum_n \delta (\omega_n - \omega') \text{i}^n J_n \left(
    \frac{\alpha a \omega'}{c_0} \right) \text{e}^{\text{i} \left( \frac{\omega'
    L}{c_0} - n \psi \right)}
  \end{array}
\end{equation}
where the final line follows from the generating function for Bessel functions {\cite{NIST:DLMF}}.  Comparison with Eq. (\ref{eq:stationary-phase-T}) shows that the single frequency transmission operator is equal to
\begin{equation}
  \begin{array}{c}
    \tau_A (\omega, \omega') = T J_{\frac{\omega' - \omega}{\Omega}} \left(
    \frac{\alpha a \omega'}{c_0} \right) \text{e}^{\text{i} \left( \frac{\omega'
    L}{c_0} - \left( \frac{\omega' - \omega}{\Omega} \right) \left( \psi -
    \frac{\pi}{2} \right) \right)} . \label{eq:bessel-transmission}
  \end{array}
\end{equation}
Substituting the above transmission function into Eq. (\ref{eq:photon-spectrum}) for the emitted photon spectrum we obtain,
\begin{equation}
  \begin{array}{ll}
    \mathcal{N} (\omega) & = \frac{1}{2 \pi T^2} \sum_{\omega_n < 0} \left(
    \frac{\omega}{- \omega_n} \right) | \tau_A (\omega, \omega_n) |^2\\
    & \\
    & = \frac{1}{2 \pi} \sum_{\omega_n < 0} \left( \frac{\omega}{- \omega_n}
    \right) \left| J_n \left( \frac{\alpha a \omega_n}{c_0} \right) \right|^2
  \end{array}
\end{equation}
which is the expression plotted in Fig. \ref{fig:photon-emission}.

\end{document}